\documentclass[showpacs,twocolumn,showkeys,preprintnumbers,amsmath,amssymb]{revtex4}
\usepackage{graphicx}
\usepackage{epsfig}% Include figure files
\usepackage{dcolumn}% Align table columns on decimal point

\begin{document}
\title{Bright entangled light from two-mode cascade laser }
\author{Eyob Alebachew Sete}
\email{eyobas@physics.tamu.edu}
%\author{K}
%\email[E-mail me at: ]{}
%\thanks{}
\affiliation{ Department of Physics  and Institute for Quantum
Studies, Texas A$\&$M University, College Station, TX 77843-4242,
USA}

\date{\today}

\begin{abstract}
We show that a two-mode three-level cascade laser driven by external
coherent fields generate intense entangled light. It turns out that
external driving fields which are at resonance with the cavity modes
substantially improves the intensity of the two-mode light in the
cavity in a region where the squeezing and entanglement is
significant making the system under consideration a viable source of
bright squeezed as well as entangled light.

\end{abstract}



\maketitle
\section{Introduction}
Generation of entanglement has recently attracted great interest as
it plays a key role in quantum information processing
\cite{1,2,3,4,5}. Particularly, much attention has been paid to
generation of continuous-variable entanglement as it might be easier
to manipulate than the discrete counterparts, quantum bits, in order
to perform quantum information processing. In general, degree of
entanglement degrades as it interacts with the environment. On the
other hand, the efficiency of quantum information processing highly
depends on the degree of entanglement. As a consequence, it is
desirable to generate strongly entangled continuous-variable states
which can survive from the environmental noise.

Schemes for generating continuous-variable states have been realized
in optical parametric oscillators \cite{6,7}. More recently,
two-mode three level cascade lasers have been proved to be a source
of macroscopic bipartite entangled states. For example, Xiong {\it
et. al.}\cite{8}, have demonstrated macroscopic entanglement in a
driven two-mode three level laser when the atoms are injected at the
lower level, applying the entanglement measure proposed by Duan {\it
et al.}\cite{9}. Tan {\it et al.}\cite{10} extended this work and
studied the generation and evolution of entanglement using the
Wigner representation. The steady state entanglement in a two-mode
three-level laser, where the atomic coherence is induced by
initially preparing atoms in coherent superposition of the top and
bottom levels has also been studied \cite{11}. The effects of
coupling this system with optical parametric oscillator \cite{12}
and two-mode squeezed vacuum reservoir \cite{13} on the steady state
two-mode squeezing as well as entanglement has thoroughly been
studied more recently. It is found that the entanglement in such
systems is directly related to the two-mode squeezing.

In this paper, we propose a scheme to produce bright entangled light
from a two-mode three-level laser driven by two coherent fields.
Three-level lasers are long known for a source of squeezed light
where the crucial role is played by atomic coherence which can be
induced by initial preparing the atoms in a coherent superposition
of the top and bottom levels \cite{14,15,16,17,18,19} or coupling
these levels by external driving fields \cite{20,21,22}. On the
other hand, the effect of driving the cavity mode by external
coherent light has been studied recently \cite{19}. In the
experimental setting, driving the OPO cavity with seed waves in a
way to improve the intensity of the signal and idler modes which
resulted from the down conversion process has been realized
\cite{23}. In this work we show that the driving coherent fields
considerably enhance the mean photon number, without affecting the
degree of entanglement obtained from the laser system, in a region
where the squeezing and entanglement is significant. This makes the
system under consideration a source of intense squeezed as well as
entangled light.

We derive the pertinent master equation in the linear and adiabatic
approximation schemes. The resulting master equation is used to
obtain equation of evolution for the first- and second- order
moments of the cavity mode operators. Using the steady state
solutions of these equations we study the two-mode steady state
squeezing in sum and difference fields. Moreover, applying the same
solutions, we analyze the entanglement properties applying the
entanglement measure proposed Duan {\it et al.} \cite{9}. Finally,
we also calculate the mean photon number of the two-mode light.

\section{Hamiltonian and Master equation}
We consider nondegenerate three-level cascade atoms injected in to a
cavity coupled to a vacuum reservoir. The atoms are initially
prepared in a coherent superposition of the top and bottom levels in
order to induce atomic coherence to the system. The atoms are
injected into the cavity at some constant rate $r_{a}$ and removed
after they decay spontaneously other than the middle and
intermediate levels. We assume that, the transition from the upper
energy level $|a\rangle$ to the intermediate level $|b\rangle$ and
from level $|b\rangle$ to the lower energy level $|c\rangle$ are
taken to be resonant with the cavity modes, whereas the transition
$|a\rangle$ $\rightarrow$ $|c\rangle$ is dipole forbidden. In turn
the cavity modes are driven by two external coherent fields having
the same frequency as the respective cavity modes.
\begin{figure}[h]
\includegraphics [height=4cm,angle=0]{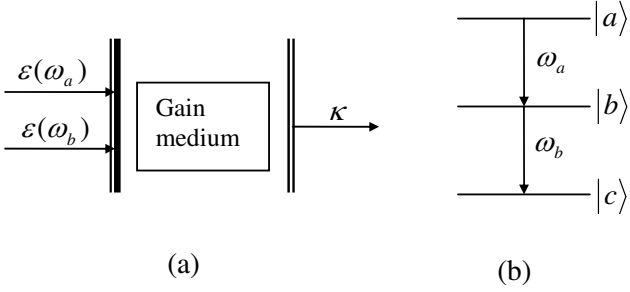}
\caption{(a) Scheme of a two-mode three-level cascade laser driven
by two coherent fields having the same amplitude, $\varepsilon$ but
different frequencies, $\omega_{a}$ and $\omega_{b}$. The gain
medium is ensembles of three-level cascade atoms. (b) Three-level
atom in a cascade configuration.}
\end{figure}

The interaction of the cavity modes with the external driving fields
and with a single three-level cascade atom is described, in the
rotating wave approximation and in the interaction picture, by the
Hamiltonian
\begin{align}\label{1}
\hat H&= i\varepsilon(\hat a^{\dagger}-\hat a+\hat b^{\dagger}-\hat
b)\notag\\
&+ig(\hat a^{\dagger}|b\rangle\langle a|+\hat b^{\dagger}|c\rangle
\langle b|-|a\rangle\langle b|\hat a-|b\rangle\langle c|\hat b),
\end{align}
where $\varepsilon$ is the amplitude of the external driving field
assumed to be the same for both fields, $g$ is the atom-cavity mode
coupling constant assumed to be the same for both transitions, $\hat
a$ and $\hat b$ are the annihilation operators for the two cavity
modes. In this work, we take the initial state of a single
three-level atom to be
\begin{equation}\label{2}
|\Psi_{A}(0)\rangle=C_{a}|a\rangle+C_{c}|c\rangle
\end{equation}
and the corresponding density operator is
\begin{equation}\label{3}
\hat\rho_{A}(0)=\rho_{aa}^{(0)}|a\rangle\langle a|
+\rho_{ac}^{(0)}|a\rangle\langle c|+\rho_{ac}^{(0)*}|c\rangle\langle
a|+\rho_{cc}^{(0)}|c\rangle\langle c|,
\end{equation}
where $\rho_{aa}^{(0)}=|C_{a}|^2$  and $\rho_{cc}^{(0)}=|C_{c}|^2$
are respectively the probabilities for the atom to be initially in
the upper and lower levels and
$\rho_{ac}^{(0)}=C_{a}C_{c}=\rho_{ac}^{(0)*}$ represents the initial
atomic coherence of the atom.

We are interested in the evolution of the cavity modes only. This
can be achieved by tracing the atom-cavity modes density operator
over atomic variables.  The time evolution of the density operator
for the cavity modes $\hat \rho(t)$ can, in general, be written as
\begin{equation}\label{4}
\frac{d}{dt}\hat \rho(t)=-iTr_{A}[\hat H, \hat \rho_{AR}(t)].
\end{equation}
Now employing Eqs. \eqref{1}-\eqref{4} and taking into the damping
of the cavity modes by the vacuum reservoir, we obtain the master
equation for the cavity modes, in the linear and adiabatic
approximation schemes, to be of the form
\begin{subequations}
\begin{align}\label{5a}
\frac{d}{dt}\hat \rho &= \varepsilon(\hat a^{\dagger}\hat\rho-
\hat \rho\hat a^{\dagger}-\hat a\hat \rho+\hat\rho\hat a)\notag\\
&+ \varepsilon(\hat b^{\dagger}\hat\rho-
\hat \rho\hat b^{\dagger}-\hat b\hat \rho+\hat\rho\hat b)\notag\\
&+ \frac{A \rho_{aa}^{(0)}}{2}(2\hat a^{\dagger}\hat \rho\hat a-\hat
a\hat a^{\dagger}\hat \rho-\hat \rho\hat a\hat
a^{\dagger})\notag\\
&+\frac{\kappa}{2}(2\hat a\hat \rho\hat a^{\dagger}-\hat
a^{\dagger}\hat a\hat\rho-\hat \rho\hat a^{\dagger}\hat a)\notag\\
&+\frac{1}{2}(A\rho_{cc}^{(0)}+\kappa)(2\hat b\hat \rho\hat
b^{\dagger}-\hat
b^{\dagger}\hat b\hat \rho-\hat\rho\hat b^{\dagger}\hat b)\notag\\
&+\frac{A\rho_{ac}^{(0)}}{2}(\hat\rho\hat a^{\dagger}\hat
b^{\dagger}+\hat a^{\dagger}\hat b^{\dagger}\hat\rho-2\hat
a^{\dagger}\hat \rho\hat
b^{\dagger})\notag\\
&+\frac{A\rho_{ca}^{(0)}}{2}(\hat\rho\hat a\hat b+\hat a\hat
b\hat\rho-2\hat b\hat\rho\hat a),
\end{align} where
\begin{equation}\label{5a}
A=2g^2r_{a}/\gamma^2
\end{equation}
\end{subequations}
is the linear gain coefficient, $\gamma$ is the spontaneous decay
rate assumed to be the same for all the three levels. $\kappa$ is
the cavity mode damping constant which we assumed it to be the same
for each cavity mode for convenience.

\section{Squeezing in the sum and difference fields}
In this section, the squeezing properties of the two-mode light in
the cavity produced by the two-mode three-level cascade laser is
analyzed. The squeezing properties a two-mode light can be
investigated by introducing sum and difference fields of the two
cavity modes and analyzing their respective quadrature variances.

We define the sum operator of the two modes as
\begin{equation}\label{6}
\hat c=\frac{1}{\sqrt{2}}(\hat a+\hat b).
\end{equation}
The corresponding quadrature operators are
\begin{equation}\label{7}
\hat c_{+}=\hat c^{\dagger}+\hat c, ~~~~~\hat c_{-}=i(\hat
c^{\dagger}-\hat c)
\end{equation}
Based on these definitions of the quadrature operators the two-mode
light represented by the sum operator is said to be in a two-mode
squeezed state provided that either of the variances of these
operators, $\Delta c_{-}^2$ or $\Delta c_{+}^2$, should be less than
that of the vacuum level, which is unity. The variances of the
quadrature operators can be expressed as
\begin{align}\label{8}
\Delta c^2_{\pm}&=\langle c_{\pm}^2\rangle-\langle c_{\pm}
\rangle^2.
\end{align}
We wish to calculate the quadrature variances at steady state. Thus,
using the steady state solutions obtained at the Appendix, the
steady state quadrature variances are found to be
\begin{align}\label{9}
\Delta c^2_{\pm}&=\frac{A^2(1-\eta^2)+(2\kappa+A+A\eta)
(2\kappa+A\eta\pm A\sqrt{1-\eta^2})
}{2(2\kappa+A\eta)(\kappa+A\eta)}.
\end{align}

Similarly we can define the difference operator of the two modes as
\begin{equation}\label{10}
\hat d=\frac{1}{\sqrt{2}}(\hat a-\hat b)
\end{equation}
with corresponding quadrature operators
\begin{equation}\label{11}
\hat d_{+}=\hat d^{\dagger}+\hat d, ~~~~~\hat d_{-}=i(\hat
d^{\dagger}-\hat d).
\end{equation}
The variances of these quadrature operators are:
\begin{align}\label{12}
\Delta d^2_{\pm}&=\langle d_{\pm}^2\rangle-\langle d_{\pm}
\rangle^2.
\end{align}
In the same way, the difference mode is said to be in a two-mode
squeezed state if either $\Delta d_{+}^2<1$ or $\Delta d_{+}^2>1$.
On account of Eqs.\eqref{10} and \eqref{11} along with the steady
state solutions at the Appendix, the quadrature variances of take
the form
\begin{align}\label{13}
\Delta d^2_{\pm}&=\frac{A^2(1-\eta^2)+(2\kappa+A+A\eta)
(2\kappa+A\eta\mp A\sqrt{1-\eta^2})
}{2(2\kappa+A\eta)(\kappa+A\eta)},
\end{align}
where $\eta=\rho_{cc}^{(0)}-\rho_{aa}^{(0)}$ which relates the
initial probabilities for an atom to be in the upper and lower
levels.
\begin{figure}
\includegraphics [height=7cm,angle=0]{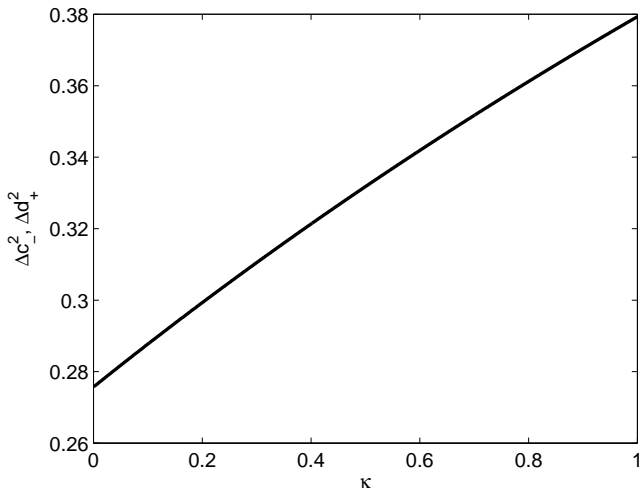}
\caption{Plot of the quadrature variances $\Delta c_{-}^2$ and
$\Delta d_{+}^2$ versus $\kappa$ for $\eta=0.1$ and A=100.}
\end{figure}
\begin{figure}
\includegraphics [height=7cm,angle=0]{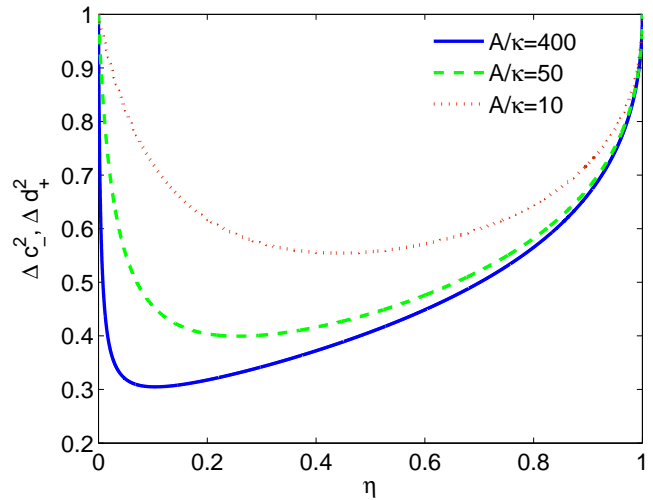}
\caption{Plots of the quadrature variance $\Delta c_{-}^2$ and
$\Delta d_{+}^2$ versus $\eta$ for $\kappa=0.15$ and for different
values of the linear gain coefficient.}
\end{figure}
According to Eq. \eqref{9} and \eqref{13}, all quadrature variances
are independent of the parameter $\varepsilon$ which represents the
driving coherent fields. This shows that the driving coherent fields
do not have any effect on the degree of squeezing of the two-mode
light. This is due to the fact that the external coherent fields do
not introduce additional coherence to the system which is believed
to be the source of squeezing in three-level cascade lasers
\cite{14,15,16,18,19}. Comparing Eqs. \eqref{9} and \eqref{13}, we
easily to see that $\Delta c^2_{+}=\Delta d^2_{-}$ and $\Delta
c^2_{-}=\Delta d^2_{+}$. It is found that exactly the same amount of
squeezing exhibited in the minus quadrature of the sum mode and in
the plus quadrature for the difference mode. It is however worth
mentioning that this statement is valid only at steady state. In the
following we explore how the two mode squeezing can be optimized by
varying the three parameters in Eqs. \eqref{9} and \eqref{13}. In
Fig. 2, we plot the $\Delta c^2_{-} $ and $\Delta d^2_{+}$ versus
cavity damping constant, $\kappa$. As can be seen from this figure
the variances increase almost linearly with $\kappa$, i.e, the
squeezing decreases linearly with $\kappa$. Obviously, $\kappa$
represents the property of one of the mirrors, port mirror, that
builds the cavity. The larger the value of $\kappa$ the more the
cavity modes susceptible to interact with the environment which
essentially degrades the degree of squeezing. We also plotted, in
Fig. 3, the variances of the squeezed quadratures versus the
parameter $\eta$ for different values of the linear gain
coefficient. It is possible to see from this plots that the two-mode
squeezing increases with the linear gain coefficient as previously
established \cite{12,13,19}. Moreover, the value of $\eta$ at which
the maximum squeezing occurs decreases to zero as the linear gain
coefficient increases. Therefore, two-mode squeezing can be
optimized by choosing small values$\eta$ (i.e, by initially prepaying slightly more atoms in the lower
level) and large values of $A/\kappa$.

\section{Entanglement between the two cavity modes}
In this section we demonstrate a continuous-variable bipartite
entanglement between the two cavity modes $a$ and $b$. Entanglement
is solely a property of quantum mechanics which is related to the
inseparability of the combined density matrix of a system into the
density matrices of the individual system. Inseparability criteria
for continuous variable bipartite states have been proposed by
various authors \cite{9,24}. In this work, we apply entanglement
criterion introduced by Duan {\it et al.} \cite {9} which is
sufficient and necessary condition for Guassian states and
sufficient, in general to detect bipartite continuous variable
entanglement. According to this criterion, a quantum state of a
system is said to be entangled if the sum of the variances of the
EPR-like quadrature operators satisfy the inequality
\begin{equation}\label{14}
\Delta u^2+\Delta v^2<2(z^2+\frac{1}{z^2}),
\end{equation}
where
\begin{equation}\label{15}
\hat u=|z|\hat x_{a}+\frac{1}{z}\hat x_{b} \end{equation}
 and
\begin{equation}\label{16}
\hat v=|z|\hat p_{a}-\frac{1}{z}\hat p_{b}
\end{equation}
with $\hat x_{a}=(\hat a^{\dagger}+\hat a)/\sqrt{2}$, $\hat
x_{b}=(\hat b^{\dagger}+\hat b)/\sqrt{2}$, $\hat p_{a}=i(\hat
a^{\dagger}-\hat a)/\sqrt{2}$, and $\hat p_{b}=i(\hat
b^{\dagger}-\hat b)/\sqrt{2}$ in which $z$ is a non-zero real
number. In the following, we choose $z=-1$ so that the upper bound
of the inequality Eq. \eqref{14} to be 2. This is not an optimal
choice in general but it is sufficient to compare with other
inequality. With $z=-1$, the quadrature operators between which we
want to demonstrate bipartite entanglement are simply the squeezed
quadrature operators for the sum and difference modes, $\hat u=\hat
c_{-}$ and $\hat v=\hat d_{+}$. We already calculated the variances
of these quadrature operators in the previous section. Thus on
account of Eqs. \eqref{9} and \eqref{13}, the sum of the variances
of $\hat u$ and $\hat v$ becomes
\begin{align}\label{17}
&\Delta u^2+\Delta v^2=2\Delta c_{-}^2=2\Delta d_{+}^2\notag\\
&=\frac{A^2(1-\eta^2)+(2\kappa+A+A\eta) (2\kappa+A\eta-
A\sqrt{1-\eta^2}) }{(2\kappa+A\eta)(\kappa+A\eta)}
\end{align}

\begin{figure}
\includegraphics [height=7cm,angle=0]{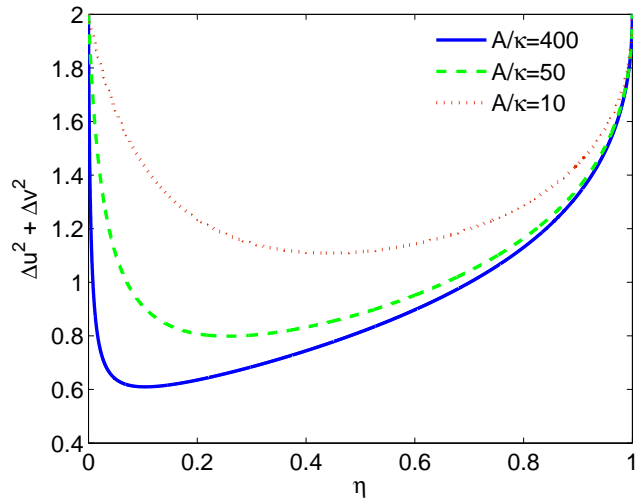}
\caption{Plots of $\Delta u^2+\Delta v^2$ of the two-mode light in
the cavity as steady state versus $\eta$ and for different values of the linear gain
coefficient.}
\end{figure}

We immediately notice that, this particular entanglement measure is
directly related the two-mode squeezing, as previously reported
elsewhere \cite{12,13}. This direct relationship shows that whenever
there is a two-mode squeezing in the system there will be
entanglement in the system as well. It also follows that the degree
of entanglement does not depend on the external driving coherent
fields. This is attributed to the fact that the coherent fields do
not introduce additional atomic coherence to the system, as the same
is true for the case of squeezing. In Fig. 4, we show results for
the sum of the squeezed quadrature variances indicating a clear
violation of the inequality and hence the entanglement between the
modes. It can easily be seen that the degree of entanglement
increases with the rate at which the atoms are injected into the
cavity, A. The entanglement however disappears at two extreme value
of $\eta$, one at $\eta=0$ which corresponds to maximum injected
atomic coherence, $\rho_{ac}^{(0)}=1/2$ and the other at $\eta=1$
corresponds to no atomic coherence, $\rho_{ac}^{(0)}=0$.

\section{Mean photon number of the cavity modes}
In order to know how intense the produced light is, we calculate the
total mean photon number of the cavity modes. In terms of the sum
and difference operators of the cavity modes, the mean photon number
can be written as
\begin{equation}\label{18}
\langle\hat N \rangle=\langle\hat c^{\dagger}\hat
c\rangle+\langle\hat d^{\dagger}\hat d\rangle.
\end{equation}
Using the definitions of the sum and difference operators( Eqs.
\eqref{6} and \eqref{10}), the mean photon number becomes
\begin{equation}\label{19}
\langle\hat N \rangle=\langle\hat a^{\dagger}\hat
a\rangle+\langle\hat b^{\dagger}\hat b\rangle.
\end{equation}
Now, with the aid of Eqs. \eqref{A19} and \eqref{A20} the mean
photon number takes the form
\begin{align}\label{20}
\langle\hat N \rangle &=\frac{A
(1-\eta)(2\kappa+A+A\eta)}{4(\kappa^2+\kappa
A\eta)}-\frac{A^3\eta(1-\eta^2)}{4(\kappa^2+\kappa
A\eta)(2\kappa+A\eta)}\notag\\
&+\frac{4\varepsilon^2[A^2(1-\sqrt{1-\eta^2})+2(\kappa^2+\kappa A
\eta)]}{(\kappa^2+\kappa A\eta)^2}.
\end{align}

In Eq. \eqref{20} the term that contain $\varepsilon$ represents the
contribution from the external driving coherent fields to the total
mean photon number. Graphically, the effect of the coherent fields
on the mean photon number of the cavity modes is shown in Figs. 5
and 6. Fig. 5, indeed, clearly indicates that the mean photon number
of the cavity modes increases with the amplitude of the coherent
driving fields. In order to clearly see by what extent the coherent
fields enhance the mean photon number over the laser system, we
plotted in Fig. 6 the mean photon number versus the parameter $\eta$
in the absence and presence of the coherent fields. It is quite
interesting to note from this figure that the coherent fields
enhance the mean photon number over the laser system by several tens
of thousands of mean photon numbers. More importantly, the increase
in the mean photon number is observed in a region where the degree
of two-mode squeezing and entanglement is significant making the
system under consideration a viable source of intense squeezed as
well as entangled light.

\begin{figure}
\includegraphics [height=7cm,angle=0]{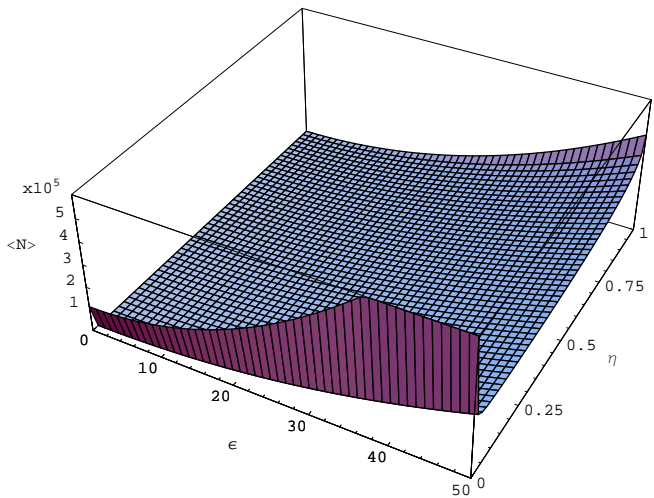}
\caption{Plot of the mean photon number versus $\varepsilon$ and
$\eta$ for $A/\kappa=100$.}
\end{figure}
\begin{figure}
\includegraphics [height=7cm,angle=0]{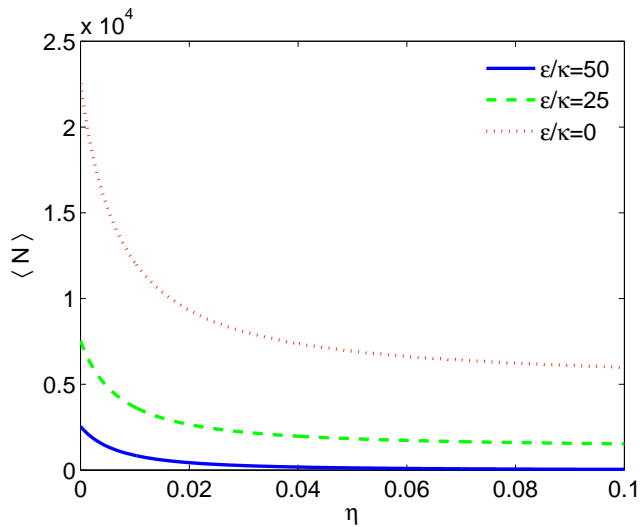}
\caption{Plots of the mean photon number versus $\eta$ for
and $A/\kappa=100$, in the absence of the coherent fields
(solid) and in the presence of the coherent fields with
$\varepsilon/\kappa=50$ (dotted).}
\end{figure}

\section{Conclusion}
We have studied the steady state two-mode squeezing and entanglement
in a two-mode three-level cascade laser driven by two external
coherent fields at resonance with the cavity modes, where the atomic
coherence is induced by initial superposition of top and bottom
levels. We have demonstrated two-mode squeezing in the sum and
difference fields with equal amount of squeezing in each field. In
addition, we have shown that the two cavity modes are strongly
entangled and the degree of entanglement is directly related to the
two-mode squeezing. We also found that the effect of the external
coherent fields is to increase the mean photon numbers considerably
in a region where the squeezing and entanglement are strong making
the system under scrutiny a viable source of a bright macroscopic
squeezed as well as entangled light. The degree of squeezing and
entanglement can be optimized by choosing small values of the cavity
damping constant, for large values of the linear gain coefficient,
and by initially preparing slightly more atoms in the lower level
than in the upper level.

\appendix
\section{Equation of evolution of cavity mode operators}
In this Appendix, we derive, applying the master equation obtained
in Sec. II, the equation of evolution for the cavity mode operators.
We also calculate the steady state solutions of the resulting
equations. The equation of evolution of an operator $\hat O$ in the
Schrodinger picture is expressible as
\begin{equation}\label{A1}
\frac{d}{dt}\langle \hat O\rangle=Tr(\frac{d\hat \rho}{dt}\hat O).
\end{equation}
Using the master equation, Eq. \eqref{5a} and Eq. \eqref{6}, we
obtained the following equations:
\begin{equation}\label{A2}
{d\over dt}\langle \hat a\rangle=-\frac{\mu_{a}}{2}\langle \hat
a\rangle -\frac{A\rho_{ac}^{(0)}}{2}\langle\hat
b^{\dagger}\rangle+\varepsilon,
\end{equation}
\begin{equation}\label{A3}
{d\over dt}\langle \hat b\rangle=-\frac{\mu_{b}}{2}\langle \hat
b\rangle +\frac{A\rho_{ac}^{(0)}}{2}\langle\hat
a^{\dagger}\rangle+\varepsilon,
\end{equation}
\begin{equation}\label{A4}
{d\over dt}\langle \hat a^2\rangle=-\mu_{a}\langle \hat a^2\rangle
-A\rho_{ac}^{(0)}\langle\hat a\hat
b^{\dagger}\rangle+2\varepsilon\langle\hat a\rangle,
\end{equation}
\begin{equation}\label{A5}
{d\over dt}\langle \hat b^{2}\rangle=-\mu_{b}\langle \hat b^2\rangle
+A\rho_{ac}^{(0)}\langle\hat a^{\dagger}\hat
b\rangle+2\varepsilon\langle\hat b\rangle,
\end{equation}
\begin{align}\label{A6}
{d\over dt}\langle \hat a^{\dagger}\hat a\rangle &=-\mu_{a}\langle
\hat a^{\dagger}\hat a\rangle-\frac{A\rho_{ac}^{(0)}}{2}(\langle
\hat a \hat b\rangle+\langle a^{\dagger}\hat
b^{\dagger}\rangle)\notag\\
&+\varepsilon(\langle\hat a^{\dagger}\rangle+\langle\hat
a\rangle)+A\rho_{aa}^{(0)},
\end{align}
\begin{align}\label{A7}
{d\over dt}\langle \hat b^{\dagger}\hat b\rangle &=-\mu_{b}\langle
\hat b^{\dagger}\hat b\rangle+\frac{A\rho_{ac}^{(0)}}{2}(\langle
\hat a \hat b\rangle+\langle \hat a^{\dagger}\hat
b^{\dagger}\rangle)\notag\\
 &+\varepsilon(\langle\hat
b^{\dagger}\rangle+\langle\hat b\rangle),
\end{align}
\begin{align}\label{A8}
{d\over dt}\langle \hat a\hat b^{\dagger}\rangle &=-\mu\langle \hat
a\hat b^{\dagger}\rangle +\frac{A\rho_{ac}^{(0)}}{2}(\langle \hat
a^2\rangle-\langle \hat b^{\dagger
2}\rangle)\notag\\
&+\varepsilon(\langle\hat a\rangle+\langle\hat b^{\dagger}\rangle),
\end{align}
\begin{align}\label{A9}
{d\over dt}\langle \hat a\hat b\rangle &=-\mu\langle \hat a\hat
b\rangle+ \frac{A\rho_{ac}^{(0)}}{2}(\langle \hat a^{\dagger}\hat
a\rangle-\langle b^{\dagger}\hat
b\rangle)\notag\\
&+\frac{1}{2}A\rho_{ac}^{(0)}+\varepsilon(\langle\hat
a\rangle+\langle\hat b\rangle),
\end{align}
in which
\begin{align}\label{A10}
&\mu_{a}=\kappa-A\rho_{aa}^{(0)},~~~~
\mu_{b}=\kappa+A\rho_{cc}^{(0)},\notag\\
& \mu=\frac{1}{2}[2\kappa+A(\rho_{cc}^{(0)}-\rho_{aa}^{(0)})].
\end{align}

It proves to be convenient to introduce a parameter that relates the
probability for an atom to be in the upper and lower levels, such
that
\begin{equation}\label{A11}
\rho_{aa}^{(0)}=\frac{1-\eta}{2},
~~~~~~~\rho_{cc}^{(0)}=\frac{1+\eta}{2}
\end{equation}
and
\begin{equation}\label{A12}
\rho_{ac}^{(0)}=\frac{1}{2}\sqrt{1-\eta^2}.
\end{equation}
Next, we calculate the steady solutions by setting the time
derivatives of the cavity mode variable to zero. By simultaneously
solving the steady state solutions, we find the following:
\begin{equation}\label{A13}
\langle\hat
a\rangle=\frac{\varepsilon[2\kappa+A(1+\eta)-A\sqrt{1-\eta^2}]}{\kappa^2+\kappa
A\eta},
\end{equation}
\begin{equation}\label{A14}
\langle\hat
b\rangle=\frac{\varepsilon[2\kappa-A(1-\eta)-A\sqrt{1-\eta^2}]}{\kappa^2+\kappa
A\eta},
\end{equation}
\begin{align}\label{A15}
\langle\hat a^2\rangle
&=\frac{4\varepsilon^2(2\kappa+A(1+\eta)-A\sqrt{1-\eta^2})}{(\kappa^2+\kappa
A\eta)(2\kappa-A(1-\eta))}\notag\\
&-\frac{\varepsilon^2A\sqrt{1-\eta^2}(2\kappa+A(1+\eta))}{(\kappa^2+\kappa
A\eta)^2}\notag\\
&-\frac{\varepsilon^2
A^3(1-\eta^2)(2-\sqrt{1-\eta^2})}{(2\kappa-A(1-\eta))(\kappa^2+\kappa
A\eta)^2},
\end{align}

\begin{align}\label{A16}
\langle\hat b^2\rangle
&=\frac{4\varepsilon^2(2\kappa-A(1-\eta)+A\sqrt{1-\eta^2})}{(\kappa^2+\kappa
A\eta)(2\kappa+A(1+\eta))}\notag\\
&+\frac{\varepsilon^2A\sqrt{1-\eta^2}(2\kappa-A(1-\eta))}{(\kappa^2+\kappa
A\eta)^2}\notag\\
&+\frac{\varepsilon^2
A^3(1-\eta^2)(2-\sqrt{1-\eta^2})}{(2\kappa+A(1+\eta))(\kappa^2+\kappa
A\eta)^2},
\end{align}
\begin{align}\label{A17}
\langle\hat a\hat b^{\dagger}\rangle
&=\frac{\varepsilon^2(2\kappa-A(1-\eta))(2\kappa+A(1+\eta))}{(\kappa^2+\kappa
A\eta)^2}\notag\\
&+\frac{\varepsilon^2A^2\sqrt{1-\eta^2}(2-\sqrt{1-\eta^2})}{(\kappa^2+\kappa
A\eta)^2},
\end{align}
\begin{align}\label{A18}
\langle\hat a\hat b\rangle
&=\frac{\kappa A\sqrt{1-\eta^2}(2k+A(1+\eta))}{4(\kappa^2+\kappa A\eta)(2\kappa+A\eta)}\notag\\
&+\frac{\varepsilon^2(2\kappa-A(1-\eta))(2\kappa+A(1+\eta))}{(\kappa^2+\kappa
A\eta)^2}\notag\\
&+\frac{\varepsilon^2A^2\sqrt{1-\eta^2}(2-\sqrt{1-\eta^2})}{(\kappa^2+\kappa
A\eta)^2},
\end{align}
\begin{align}\label{A19}
\langle\hat a^{\dagger}\hat a\rangle &=\frac{
A(1-\eta)(2\kappa+A+A\eta)}{4(\kappa^2+\kappa A
\eta)}-\frac{(\kappa+A\eta)
A^2(1-\eta^2)}{4(2\kappa+A\eta)(\kappa^2+\kappa
A\eta)}\notag\\
&+\frac{4\varepsilon^2(2\kappa+A(1+\eta)-A\sqrt{1-\eta^2})}{(2\kappa-A(1-\eta))(\kappa^2+\kappa
A\eta)}\notag\\
&-\frac{\varepsilon^2
A\sqrt{1-\eta^2}(2k+A(1+\eta))}{(\kappa^2+\kappa A\eta)^2}\notag\\
&-\frac{\varepsilon^2A^3(1-\eta^2)(2-\sqrt{1-\eta^2})}
{(2\kappa-A(1-\eta))(\kappa^2+\kappa A\eta)^2},
\end{align}
\begin{align}\label{A20}
\langle\hat b^{\dagger}\hat b\rangle &=\frac{\kappa
A^2(1-\eta^2))}{4(2\kappa+A\eta)(\kappa^2+\kappa A\eta)}\notag\\
&+\frac{4\varepsilon^2(2\kappa-A(1-\eta)+A\sqrt{1-\eta^2})}{(2\kappa+A(1+\eta))(\kappa^2+\kappa
A\eta)}\notag\\
 &+\frac{\varepsilon^2
A\sqrt{1-\eta^2}(2k-A(1-\eta))}{(\kappa^2+\kappa A\eta)^2}\notag\\
&+\frac{\varepsilon^2A^3(1-\eta^2)(2-\sqrt{1-\eta^2})}
{(2\kappa+A(1+\eta))(\kappa^2+\kappa A\eta)^2}.
\end{align}
It is worth mentioning that these solutions are obtained provided
that $\eta>0$.


\begin{thebibliography}{1}
%%%%
\bibitem{1} J.-M. Liu, B.-S. Shi, X.-F. Fan, J. Li
and G.-C. Guo, J. Opt. B: Quantum Semiclass. Opt. 3 (2001) 189–193
\bibitem{2} S. L. Braunstein and H. J. Kimble,  Phys. Rev. A {\bf
61}, (2000) 042302.
\bibitem{3} S. Lloyd and S. L. Braunstein,  Phys. Rev. Lett. {\bf 82}, (1999) 1784.
\bibitem{4} S. L. Braunstein, Nature {\bf 394}, (1998) 47.
\bibitem{5} T. C. Ralph, Phys.Rev. A {\bf 61}, (2000) 010302.
\bibitem{6} M. D. Reid and P. D. Drummond, Phys. Rev. Lett. {\bf 60},
(1988) 2731.
\bibitem{7} Y. Zhang, H. Wang, X. Li, J. Jing, C. Xie, and K. Peng, Phys.
Rev. A 62, (2000) 023813.
\bibitem{8} H. Xiong, M. O. Scully, and M. S. Zubairy, Phys. Rev.
Lett. {\bf 94}, (2005) 023601.
\bibitem{9}L.-M. Duan, G. Giedke, J. I. Cirac, and P. Zoller, Phys.
Rev. Lett. {\bf 84}, (2000) 2722.
\bibitem{10}H.-T. Tan, S.-Y. Zhu, and M. S. Zubairy, Phys. Rev. A
{\bf 72}, 022305 (2005).
\bibitem{11} S. Tesfa, Phys. Rev. A {\bf 74}, (2006) 043816.
\bibitem{12} E. Alebachew, Phys. Rev. A, {\bf 76} (2007) 023808.
\bibitem{13} E. Alebachew, Opt. Commun. {\bf 280} (2007) 133.
\bibitem{14}M. O. Scully, K. Wodkiewicz, M. S. Zubairy, J. Bergou, N. Lu, and
J. Meyer ter Vehn, Phys. Rev Lett. {\bf 60}, (1988) 1832.
\bibitem{15}K. Fesseha, Phys. Rev. A {\bf 63}, (2001) 033811.
\bibitem{16}N. Lu, F. X. Zhao, and J. Bergou, Phys. Rev. A {\bf
39}, (1989) 5189.
\bibitem{17}N. Lu and S. Y. Zhu, Phys. Rev. A {\bf 40}, (1989) 5735.
\bibitem{18}C. A. Blockley and D. F. Walls, Phys. Rev. A {\bf 43}, (1991) 5049.
\bibitem{19}E. Alebachew, Opt. Commun. {\bf 273}, (2007) 488.
\bibitem{20}N. A. Ansari, J. Gea-Banacloche, and M. S. Zubairy,
Phys. Rev. A {\bf  41}, (1990) 5179.
\bibitem{21}N. A. Ansari, Phys. Rev. A {\bf 48}, (1993) 4686.
\bibitem{22}E. Alebachew, K. Fesseha, Opt. Commun. {\bf 265}, (2006) 314.
\bibitem{23}V. D'Auria, S. Fornaro, A. Porzio, E. A. Sete, and S.
Solimeno, Appl. Phys. B, {\bf 91},(2008) 309.
\bibitem{24} R. Simon, Phys. Rev. Lett. {\bf 84}, 2726 (2000).
\end{thebibliography}
\end{document}